# Implementation of continuous variable quantum cryptography in optical fibres using a go-&-return configuration

Matthieu Legré, Hugo Zbinden, Nicolas Gisin
Group of Applied Physics, University of Geneva, 1211 Geneva 4, Switzerland
E-mail:Matthieu.legre@physics.unige.ch

*Abstract: We demonstrate an implementation of quantum key distribution with continuous variables based on a go-&-return configuration over distances up to 14km. This configuration leads to self-compensation of polarisation and phase fluctuations. We observe a high degree of stability of our set-up over many hours.*

## Introduction

It has been demonstrated theoretically that several protocols of quantum key distribution (QKD) based on continuous variables allow one to achieve a large quantum bit rate. Some implementations of those protocols have already shown the potential of this technique[1, 2]. However, the efficiency of those protocols is very sensitive to losses and the reconciliation protocols aren't efficient enough for the moment to obtain the secret bit rates predicted by the Shannon theory. In [1], it has been shown theoretically that, considering coherent states and reverse reconciliation, QKD secure against individual attacks can be achieved for any arbitrary loss. In practice, the system is limited to losses of roughly 4-5dB.

Some recent studies have demonstrated that continuous variable QKD using coherent states and homodyne detection can be unconditionally secure if the loss is smaller than 1.9dB [3, 4]. So, quantum cryptography based on continuous variable has the advantage of having a high secret bit potential per pulse, but on the other hand it works only for small losses and is limited by the existing classical secret bit distillation protocols.

Our objective is to demonstrate the feasibility of QKD with continuous variables over a distance of a few kilometres by using low loss optical fibres (~0.2dB/km). 10 kilometres corresponds to 2dB-loss, which is a reasonable loss value for secure key exchange with continuous variables, and this distance is long enough to allow the implementation of quantum cryptography in local networks. However, the use of fibre adds problems such as polarisation and phase fluctuations with time. These problems have already been solved for discrete variable quantum cryptography with the *Plug-&-Play* configuration [5]. We apply this *go-&-return* configuration to continuous variable quantum cryptography.

In the first section, we present our pulsed homodyne detector at 1.55μm with pigtailed components. Then, in section 2, we describe our experimental set-up. We show some experimental results on the homodyne detection of coherent states, and on the stability of our experiment in the third section. We conclude with some discussions about possible improvements to our set-up.

**Pulsed homodyne detection with optical fibres**

The detection technique used for continuous variable QKD is pulsed homodyne detection. First, we describe this technique briefly, and then we show the technical solutions that have been adopted for working with fibres.

The principle of the standard homodyne detection is shown in figure 1.a. This detection allows one to extract the amplitude and phase of a light beam, called the signal. The phase of the signal is defined with respect to the phase of a second light beam, called the local oscillator (LO). As can be seen in figure 1.a, the idea of homodyne detection is to mix the two light beams at a 50/50 beam-splitter and to measure the difference of intensities between the two outputs of the beam-splitter. The result of homodyne detection is proportional to the part of the amplitude of the signal that is in quadrature with the local oscillator ($\delta i = 2q\sqrt{I_{LO}}$, where $I_{LO}$ is the intensity of the LO and $q$ is the part of the amplitude of the signal that is in quadrature with the local oscillator).

Pulsed homodyne detection is a bit different from standard homodyne detection. "Pulsed" means that the duration of the optical pulse is very short compared to the response time of the detector. This means that the shape of the electrical signal output is independent of the input light, only the amplitude of the electrical signal changes with the amplitude and the phase of the input optical pulse. In order to have a large repetition rate, the temporal duration of the electrical output pulse must be small. But the electrical noise of the detector increases with the spectral bandwidth of the electrical amplifier. Therefore, to make a homodyne detector limited by the shot-noise, a trade-off has to be found between the duration of electrical pulse and the electrical noise.

For the electrical amplifier, we have chosen to use a circuit composed of a charge sensitive amplifier and a pulse shaper. The electronic amplifiers are AMPTEK A250 and A275 as used in [6]. These electrical circuits and components allow us to get a detection system limited by the shot-noise with an electrical pulse duration of ~5μs. This provides for a maximal frequency repetition rate of ~200kHz.

The optical part of the homodyne detection is implemented with fibred components. To be shot-noise limited, the homodyne detection has to be balanced. For example, if we consider a LO intensity of $10^8$ photons per pulse, the precision on the ratio 50/50 has to be $4*10^{-5}$ [7], expressed in term of dB this means $-3\pm0.0003$dB. We balance the homodyne detection by inserting extra losses in one of the two arms of the beam splitter. With fibres, it is very simple to add a small loss by winding the fibre in small loops. Then, to balance our detection, we use a standard fibred 50/50 coupler and we loop one of the two output fibres as shown in figure 1.b. This technique works very well generally, but for a large local oscillator ($>10^8$ photons/pulse) we find that the stability of the balance is not sufficient. This comes from the fact that polarisation fluctuates in a fibre, and that every fibred components have polarisation dependent loss (PDL). The combination of the PDL of our coupler and of the polarisation fluctuations leads to the instability of our detection system. Notice that a coupler with the smallest PDL achievable has a PDL value of ~0.01dB, which is more than one order of magnitude higher than the precision needed to balance the detection for a LO intensity of $10^8$ photons per pulse. Therefore, we have to stabilize the polarisation before the coupler, as well as the temperature of our device. In order to control the polarisation without perturbing the temperature stability, we use computer driven polarisation controllers (General Photonics Corp.).

In order to check the quality of the balance of our homodyne detection, we measure the noise for several intensities of our local oscillator. If the detection system is limited by the shot-noise, the noise variance should vary linearly compared to the intensity of the LO. The results obtained by adjusting the balance with the method explained above are shown in figure 2. For clarity, we fit the data points with a linear function given by equation (1) $V = V_{electr} + aI_{OL}$, where $V_{electr}$ is a constant corresponding to the noise variance of the electronics, and *a* is a constant linked to the gain of the electrical amplifier and to the efficiency of the photodiodes. As can be seen for intensities of the local oscillator smaller than $10^6$ photons per pulse, the system is limited by the noise of the electronic amplification. But, for stronger intensities of the LO, the noise variance grows linearly with them. We can conclude that our method of balancing the homodyne detection is stable enough for the LO intensities we are going to use in the quantum cryptography system ($I_{LO}$ between $10^7$ and $10^8$ photons per pulse). Notice that for this range of intensities, the weight of the electrical noise compared to the shot-noise is between 1 and 10%.

**Experimental Set-up**

The experimental scheme is depicted in figure 3. It is very similar to the standard *Plug-&-Play* system [5] with the light propagating back and forth between Bob and Alice. When Bob sends an optical pulse to Alice, it is split into two pulses. In Bob's device, one pulse takes the short path and the second pulse takes the long one. The two pulses travel one after the other in the same fibre when they leave Bob, so there is no phase fluctuation between the two optical pulses (signal and local oscillator). Alice chooses what coherent state she sends back to Bob by modulating the amplitude and the phase of the optical signal pulse. This modulation is made with a dual electrode intensity modulator, which allows us to modulate amplitude and phase at the same time. Then, the light is reflected by Alice's Faraday mirror [8]. The effect of this mirror is that the light comes back to Bob with an orthogonal polarisation compared to its polarisation when it left. Therefore, when they come back to Bob, the two pulses reverse their paths. This shows that there is a global auto-compensation of the phase of the system and of the polarisation transformations in the fibre link. The phase is compensated for fluctuations that can be considered as constant during the separation duration between the two optical pulses (~50ns). The polarisation fluctuations are compensated if they vary slower than the propagation time of the light in the fibre (1km is equivalent to 2*5µs).

There are essentially three main differences between this scheme and the standard *Plug-&-Play* configuration. Firstly, in our system we have two pulses with significantly different intensities. The optical signal is between 0 and ~100 photons per pulse, whereas the local oscillator is between $10^7$ and $10^8$ photons per pulse. Notice that the difference is about 60-70dB. The two pulses take the same path but in opposite directions in Bob's device. Therefore, we have chosen to use an optical isolator as this component has ~60dB loss in one direction and only 1dB in the other direction. Secondly, the quality of homodyne detection is very sensitive to the loss in the signal path. We have replaced the circulator, normally used in *Plug-&-Play* systems, by a 98/2 coupler. This component has a loss of only 0.1dB, whereas a circulator has a loss of around 1dB. Furthermore, as we can see in figure 3, all the optical components of Bob's device are on the path of the LO when light comes back to Bob. The loss of the signal in Bob's device equals 2.4dB (see figure 3) and can be further reduced with better quantum efficiency of the photodiodes. Unfortunately, the quantum efficiency of InGaAs photodiodes is intrinsically quite poor, so this value of 2.4dB can't be reduced significantly. The last thing is that the ratio in intensities between signal and LO is 60dB so any reflections of the LO can have a larger intensity than the signal. In the case of our cryptographic scheme, there are many optical components though most of them aren't in the signal path. As we will see, this reduces strongly the degradation due to the reflections of the LO. Indeed, what is important is when the reflected light comes back along the signal path, so the back-reflections can be analysed as follows. There is a large reflection of the LO on the detectors. This light can be reflected back on the 50/50 coupler, it isn't important because it occurs after the mixing between the signal and the LO. But there is a reflection on the PBS also, and this light coming back is similar to the optical signal. The temporal separation between the optical signal and this reflected light frees us from degradation of the electrical signal.

**Experimental results.**

In the following experiments, we typically operated at a frequency repetition rate of 50kHz, with 50ns-duration optical pulses. Firstly, we measured some coherent states sent by Alice with Bob's homodyne detector, and then we tested the time stability of our experimental set-up.

In the first experiment, we used two different fibre spools measuring 20m and 14km. We made the same measurement with each spool. Alice applies an approximately $2\pi$ phase change in 17 steps to the coherent state which she sends. Bob always measures the same quadrature. They repeat the same operations for coherent states of different amplitude. A typical result is shown in figure 4. These graphs correspond to a measurement made with the fibre spool of 20m-length. Intensities of coherent states were between 9 and 0.01 photons per pulse. The graphs correspond to the result of the homodyne detection as a function of the voltage applied to the phase modulator, which is equivalent to the phase difference between the signal and the LO. As we can see, each graph is a cosine curve as expected by the theory. The curves have been slightly translated one compared to another in order to better distinguish them.

The same kind of measurement was made with the spool of 14km-length. Graphs, similar to the one in figure 4, were obtained. Notice that for the 14km-length spool, a repetition rate of only 6.7kHz has been used. Indeed, when the experiment was performed with a repetition rate of 50kHz, the noise of the detection strongly increased. This is due to the presence of two sets of two optical pulses (LO and signal) in Bob's device at the same time. One set comes back from Alice and the other one comes from Bob's laser. The back-reflections and Rayleigh backscattering of the set of pulses sent by Bob significantly increase the noise of Bob's homodyne measurement of the set coming back from Alice. For the moment, the only solution we have found to avoid this problem is to decrease the frequency rate until only one set of optical pulses is in the set-up at any given time. As demonstrated in [5], another solution would be to use a delay line in Alice's device and trains of pulses.

In order to control the quality of our homodyne detection, we can extract from those graphs the photon number computed with the homodyne detection results and compare this number with the photon number measured with a power-meter. To evaluate the photon number with power-meter, we first measure the extinction ratio between the local oscillator and the signal when no modulation is applied to the signal. This extinction ratio equals the difference of attenuation one has to apply to the signal and the LO to obtain the same number of counts on a photon counting detector. Then, we calibrate our amplitude and phase modulator with intense light. By using these two calibrations and a measurement of the intensity of the LO, one can extract the value of the mean number of photons for the coherent states sent by Alice. The results obtained are shown in table 1. The photon numbers given in the table correspond to those arriving at the 50/50 coupler in Bob's device. As we can see in the table, the computed errors are relatively small (~5%) for a low amplitude modulation. For both fibre lengths, this error becomes larger when Alice applies a strong attenuation on the optical signal. We can see that a precision of 15% is obtained with the 20m-length fibre for an intensity of ~1 photon per pulse, whereas the precision equals ~5% for similar intensity in the case of the 14km-length fibre. This means that the large errors, obtained with the 20m-length fibre for intensities smaller than 3 photons per pulse, aren't due to the homodyne detection but they come from the precision of the amplitude modulation. This can be understood by considering the precision we have on the position of the modulation gate. The optical pulse duration is 50ns and the modulation gate duration is ~70ns. This gate is not perfectly flat. Therefore, we are very sensitive to small shifts of this gate relative to the optical pulse for the case of strong amplitude modulation, whereas this effect is almost negligible for low amplitude modulation. It is very simple to adjust the two gates with bright light for the calibration of the amplitude modulator, but this adjustment is less precise when we make it with weak coherent states.

So we can conclude that our experimental set-up is capable of measuring coherent states up to distances of at least ~14km. The error computed between the values obtained with the homodyne detection and the one measured with standard detection are reasonably small. Essentially, this error comes from calibration and timing precision of the gate modulation.

To further characterise our set-up, we measure the stability of the system over time. We repeat the same measurement five times, waiting 1 hour between each measurement (one measurement takes less than 5 minutes), and no alignment is made during this entire 4 hours period. We use a fibre spool of 1km-length for this measurement. We proceed for each measurement as follows. Alice sends a series of coherent states with different intensities. She changes the phase of each state on a ~$2\pi$-range in 17 steps, so that this state is rotated in the phase plan. Bob measures both quadratures of each state 4000 times. Then, Bob can compare the evolution of the states for the five measurements. The results of Bob's measurements for a coherent state with an intensity of 21 photons per pulse are shown in figure 5. Each data point corresponds to the mean value of an acquisition of 4000 data points. As we can see in the figure, the curves are quite well superposed. This means that our set-up is stable in time with a fibre spool of 1km-length between Alice and Bob. Notice that no particular stabilisation is used except the temperature controls of Alice and Bob's devices, furthermore this experiment was made on a normal table on a parquet floor where people were walking close by. We calculate the reproducibility of the measurement by computing the standard deviation of the five measured values for a given state. The computed mean reproducibility equals 0.25. This can be compared with the standard deviation of the homodyne measurement noise, which is ~1.1 and with the standard deviation of the shot-noise (noise of a measurement with a perfect homodyne detection and a fibre with no loss), which is 0.707 with our convention. Then our reproducibility is significantly smaller than the homodyne measurement noise, even if we consider a perfect measurement. Of course, if the set-up had been perfectly stable, we would have measured a standard deviation of $1.1/\sqrt{4000} = 0.02$. The difference between this expected value and the measured value can be partly explained by changes of our laser power by more than 7% during the experiment. With this value of 0.25 and an amplitude value of ~6.3, we can calculate an error for the determination of the state's amplitude of ~4%. Notice that the precision on the intensity is 2 times larger than that for the amplitude. Therefore, we can maintain a precision on the intensity of 8% during 5 hours. Similar errors have been computed for signal intensities between 11 and 1 photons per pulse. This value is larger than the absolute precision we have on one measurement, but this increase can be explained by the fluctuations of the laser power.

**Conclusion and discussions**

We have implemented a pulsed homodyne detection system working with fibred components. Our device works for a large range of LO intensities, and its maximal repetition rate is currently ~200kHz. Its characteristics are sufficient for a quantum cryptography implementation based on continuous variables. To implement a continuous variable quantum key exchange in optical fibres, we have chosen to work with a *Plug-&-Play* configuration to avoid the problems of phase and polarisation fluctuations. This configuration allows us to work with a fibre link measuring several kilometres. We have shown that the precision on the determination of the intensity of a coherent state is around 5%. Furthermore, the stability of our experiment is good enough to maintain this precision during measurements of several hours. This highlights the efficiency of the stabilization attainable with the *Plug-&-Play* configuration.

Even though we didn't make a quantum key exchange, we can compute the secret key bit rate we can expect with the results shown before. We will use the formulas given in [1], so we will consider the case of reverse reconciliation secure against individual attacks. Furthermore, we will consider the "realistic" case [1], Eve can't get any information from the noise of the homodyne detection which occurs in Bob's device. The secret key bit rate will be computed for the fibre spools of 1km and 14km. These two fibres have 0.88dB and 3.1dB of loss respectively. Notice that the loss of the 1km-length fibre is larger than the expected value (1*0.2=0.2dB), this is due to a bad connector. This excess loss isn't very important for the study of the stability of our set-up. Because of the maximal power of our laser, Alice can send a maximal intensity of ~30 photons per pulse when we use the short fibre and ~20 photons per pulse when we use the other fibre. Considering those values and a reverse reconciliation protocol with 100% efficiency, we can expect a secret key bit rate of 0.63 bit/pulse with the 1km-length fibre and 0.18 bit/pulse with the 14km-length fibre. With our current repetition rate, we therefore obtain a value of 31.5 kbit/s and 1.2 kbit/s respectively for the two distances.

Those secret key bit rate values aren't very high but the main goal of our experiment was to evaluate the feasibility of the implementation of a continuous variable quantum cryptography device at 1.55μm with optical fibres. Our results show that this implementation is realistic. The first improvement one could make to our set-up is to increase the frequency repetition rate. As demonstrated in [9, 10], the pulsed homodyne detection can be improved to work at a frequency of 1MHz or even more. This would produce an increase in our secret key bit rate of more than one order of magnitude. Secondly, our set-up, with its current configuration, isn't robust against 'Trojan horse' attacks. So another improvement should be to make it free from these kinds of attacks. With our configuration, Alice can't check if the spy Eve has modified the signal pulse arriving from Bob because of the large difference in intensities between the signal and the local oscillator. If we introduce the strong attenuation in Alice's device, the two pulses arrive at Alice's device with the same intensity. Then Alice can measure the intensity of the signal pulse in order to check if Eve tries to make a 'Trojan horse' attack. This strong attenuation can be applied with an acousto-optic modulator as demonstrated in [11].

This experiment shows the feasibility of the implementation of QKD with continuous variables in optical fibres. The stability over many hours of the system, and of the homodyne detection particularly, allows us to consider the use of this kind of system in optical fibre local networks.

***Acknowledgements***: *We thank M. Wegmüller for discussions. Financial support from the Swiss OFES in the frame of the SECOQ project, EXFO Inc (Vanier, Canada), and the Swiss NCCR "Quantum photonics" are acknowledged.*

# Captions

Figure 1:   a. Schematic of the homodyne detection principle.
            b. The key components for a fibred homodyne detector

Figure 2: Detector noise measurement. The variance changes linearly compared to the intensity of the local oscillator; the detector is shot-noise limited over the intensity range $10^6$-$10^8$ photons per pulse. Circle: data point measurement, line: fit with the equation (1).

Figure 3: Set-up of the continuous variables quantum cryptography experiment. PM: phase modulator; ISOL: isolator; PBS: polarisation beam splitter; AM: amplitude modulator; FM: Faraday mirror, L: loss.

Figure 4: Measurement of coherent states. Value of the quadrature measured by Bob when Alice changes the phase of her coherent state. The different curves correspond to different intensities of the coherent states sent by Alice.

Figure 5: Measurement of the time stability of our experimental set-up. Each curve corresponds to the evolution of a coherent state of 21 photons when Alice changes its phase. Each measurement is taken at regular time intervals of 1 hour.

Table 1: Table of comparisons between the photon number computed with the homodyne measurement and the one measured directly. The two left columns correspond to a measurement with the 20m-length fibre and the two right columns correspond to a measurement with the 14km-length fibre.

figure 1a and 1b

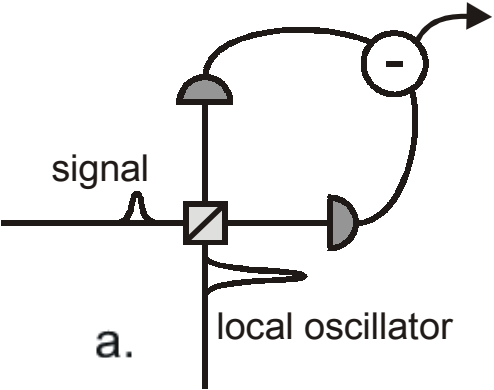 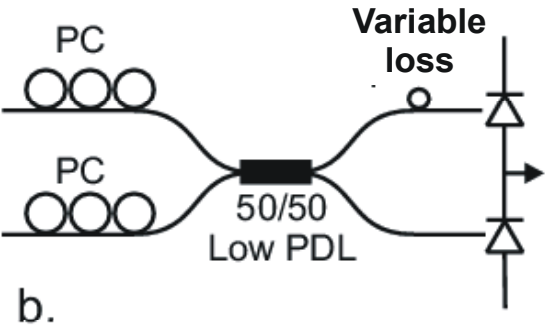

a. 　　　　　　　　　　　　b.

figure2

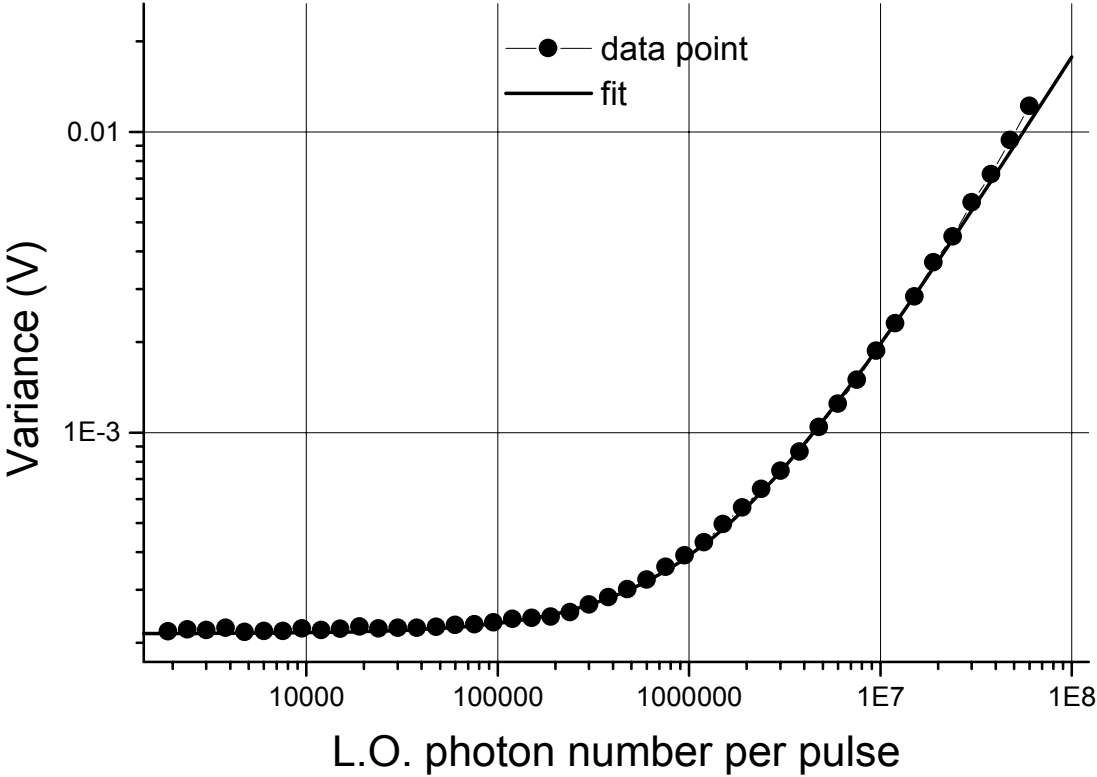

figure 3

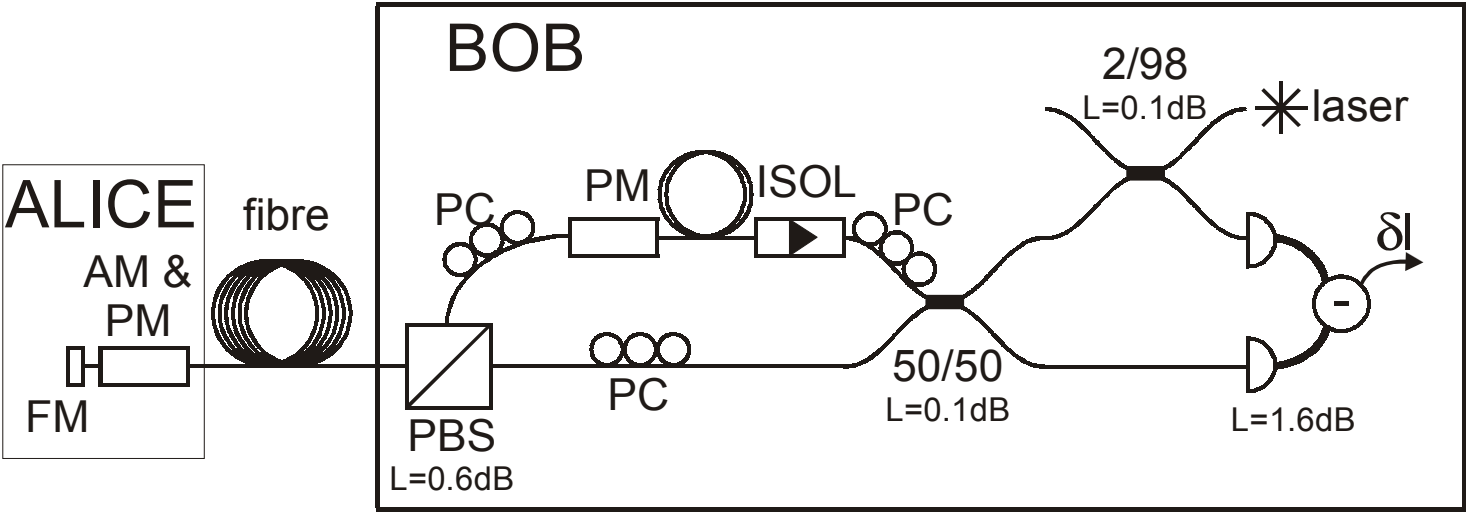

figure 4

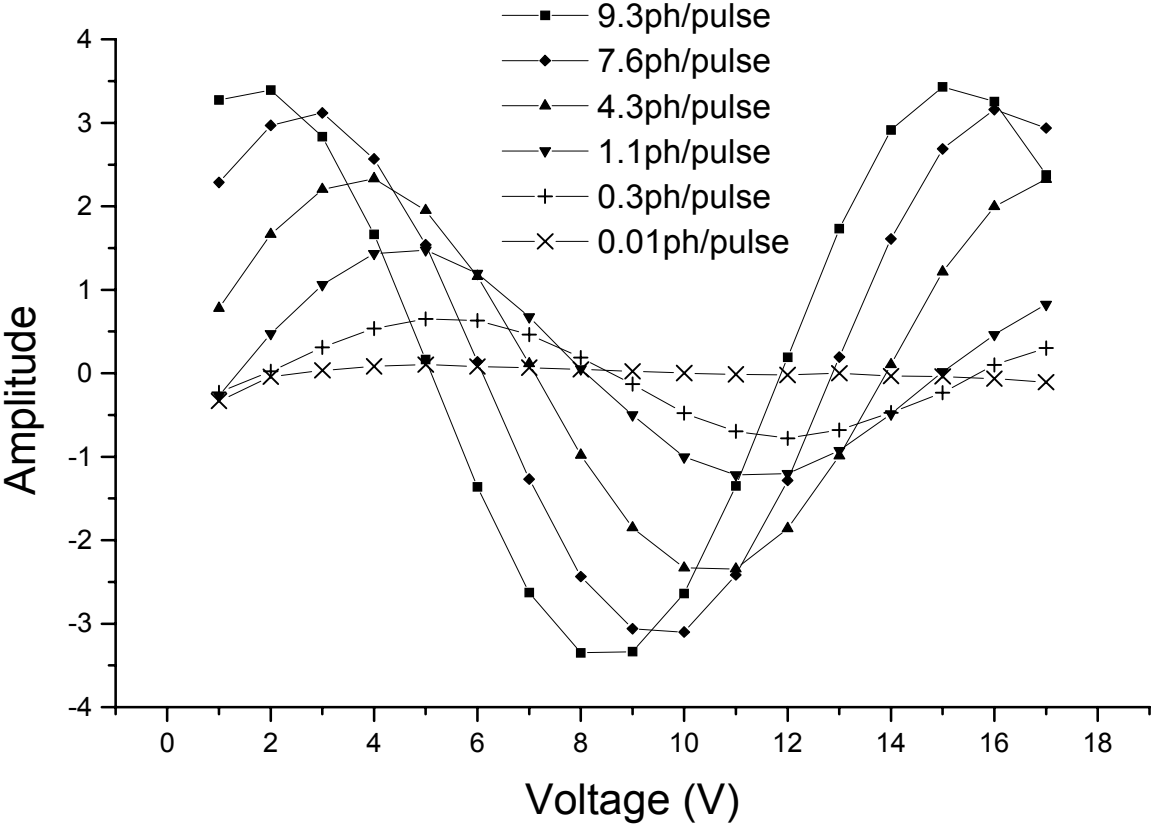

figure5

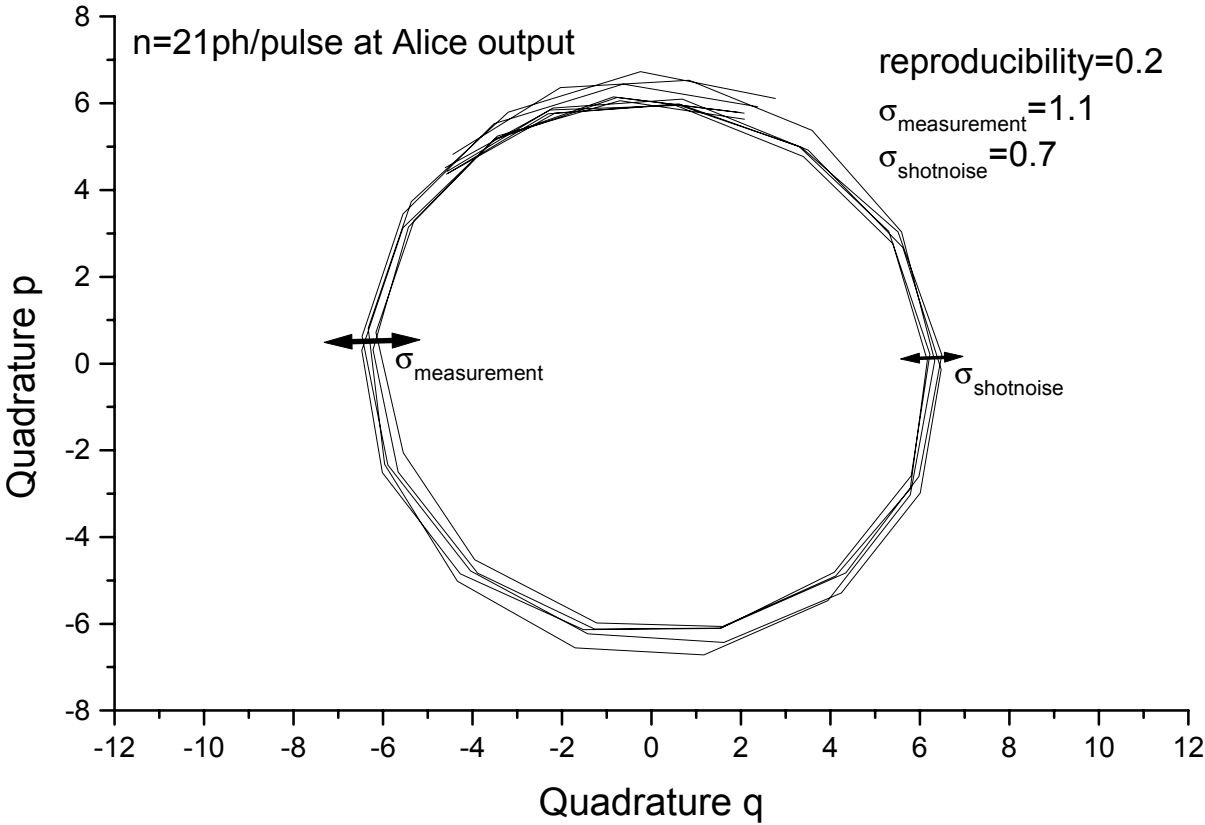

table 1

| 20m-length fibre | | | 14km-length fibre | | |
|---|---|---|---|---|---|
| Photon number standard detection | Photon number homodyne detection | error % | Photon number standard detection | Photon number homodyne detection | error % |
| 9.3 | 9.1 | 2.2 | 2.2 | 2.3 | 4.4 |
| 8.77 | 8.75 | 0.2 | 2.08 | 2.21 | 6.1 |
| 7.62 | 7.65 | 0.4 | 1.8 | 1.9 | 5.4 |
| 6 | 6.19 | 3.1 | 1.42 | 1.6 | 11.9 |
| 4.27 | 4.44 | 3.9 | 1.01 | 1.06 | 4.8 |
| 2.52 | 2.72 | 7.6 | 0.6 | 0.69 | 14 |
| 1.11 | 1.3 | 15.8 | 0.26 | 0.33 | 23.7 |
| 0.28 | 0.38 | 30.3 | 0.07 | 0.11 | 44.4 |